\documentclass[a4paper]{raa}
\usepackage{graphicx,times}
\usepackage{natbib}
\usepackage{amssymb,amsmath}
\usepackage{enumerate}
\usepackage{multirow}
\bibpunct{(}{)}{;}{a}{}{,}

\usepackage[a4paper=true,pagebackref=true]{hyperref}
\hypersetup{pdftitle = The title of my PDF, pdfauthor = My name, pdfsubject= The subject, pdfkeywords = keyword1 keyword2 keyword3}
\hypersetup{colorlinks = true, linkcolor = green, anchorcolor = red, citecolor = cyan, filecolor = red, pagecolor = red, urlcolor = red}

\begin{document}

   \title{Propagating Slow Sausage Waves in a Sunspot Observed by the New Vacuum Solar Telescope$^*$
   \footnotetext{$*$ Supported by the National Natural Science Foundation of China.}
\footnotetext{\textdagger \ Corresponding author}
}

 \volnopage{ {\bf 20XX} Vol.\ {\bf X} No. {\bf XX}, 000--000}
   \setcounter{page}{1}

   \author{Song Feng\inst{1,2}, Zheng Deng\inst{1}, Ding Yuan\inst{3}{\textsuperscript{\textdagger}}, Zhi Xu\inst{4}, Xiao Yang\inst{2}}

\institute{ Faculty of Information Engineering and Automation, Kunming University of Science and Technology, Kunming 650500, China;
	{\it feng.song@kust.edu.cn}\\
	\and
	Key Laboratory of Solar Activity, National Astronomical Observatories, Chinese Academy of Sciences, Beijing 100012, China
	\and
	Institute of Space Science and Applied Technology, Harbin Institute of Technology, Shenzhen, Guangdong 518055, China; {\it yuanding@hit.edu.cn}
	\and
	Yunnan Astronomical Observatory, Chinese Academy of Sciences, Kunmin 650011, China\\
	\vs \no
	{\small Received 20XX Month Day; accepted 20XX Month Day}
}

\abstract{
A sunspot is an ideal waveguide for a variety of magnetohydrodynamic waves, which carry a significant amount of energy to the upper atmosphere and could be used as a tool to probe magnetic and thermal structure of a sunspot. In this study, we used the New Vacuum Solar Telescope and took high-resolution image sequences simultaneously in both TiO (7058$\pm$10 \AA) and H$_\alpha$~(6562$\pm$2.5~\AA) bandpasses. We extracted the area and total emission intensity variations of sunspot umbra and analyzed the signals with synchrosqueezing transform. We found that the area and emission intensity varied with both three and five minute periodicity. Moreover, the area and intensity oscillated in phase with each other,  this fact hold in both TiO and H$_\alpha$ data. We interpret this oscillatory signal as propagating slow sausage wave. The propagation speed is estimated at about 8 km$\cdot$s$^{-1}$.  We infer that this sunspot's umbra could have temperature as low as 2800--3500 K. 
}

   \authorrunning{\textit{S. Feng et al.}: Propagating Slow Sausage Waves in a Sunspot Observed by NVST}            
   \titlerunning{\textit{S. Feng et al.}: Propagating Slow Sausage Waves in a Sunspot Observed by NVST}  
   \maketitle

%
\section{Introduction}
The solar atmosphere is a highly dynamic and magnetized plasma. Many solar activities and phenomena are observed in the solar atmosphere. 
Magnetohydrodynamic (MHD) waves are commonly found in the magnetized plasma and can be observed in a variety of magnetic structures, such as sunspots \citep{2013ApJ...779..168J,2016ApJ...816...30S,2016A&A...594A.101Y,2020raa,2013SoPh..282..405B}, pores \citep{2014A&A...563A..12D,2015ApJ...806..132G,2015A&A...579A..73M,2016ApJ...817...44F}, 
and coronal loops \citep{2009A&A...503..569I,2012ApJ...761..134N,2015ApJ...807...98Y,2016ApJ...823L..16T}.   

MHD waves are usually modified by magnetic field. They could perturb the plasma density and magnetic field, therefore, they are further classified as Alfv\'en waves \citep{2013ApJ...768..191G,2014MNRAS.445...49B}, slow and fast magnetoacoustic waves \citep{2009A&A...503..569I,2014ApJ...781...92V,2014A&A...568A..31L,2016ApJ...833...51Y,2018ApJ...857...28K,2015SoPh..290.2231C,2018ApJ...868....5C}.   
Linear Alfv\'en waves can propagate along magnetic fields without perturbing the plasma pressure and density, whereas slow and fast magnetoacoustic waves are compressible waves associated with perturbations in the plasma and magnetic pressure. 

Sausage modes are compressible magnetoacoustic waves.  
It is featured by axially symmetric perturbations in magnetic flux tubes. 
It attracts interest because of its potential role in chromoshperic and coronal heating. 
They could be classified into slow and fast modes based on their axial phase speeds \citep{1983SoPh...88..179E}. 
Another main difference between two sausage-mode forms is the phase relationship between their cross-section area and intensity variations.  In case of fast sausage mode, these two values oscillate out-of phase, whereas for slow sausage modes, they are in phase 
\citep{2013A&A...555A..75M,2014A&A...563A..12D}. Investigating sausage waves and their propagation in the lower solar atmosphere is important for understanding the wave energy propagation and release in the entire solar atmosphere. 

In the coronal context, it has been a common practice to invoke sausage modes to account for quasi-periodic pulsations in solar flare light curves with quasi-periods of the order of seconds (for recent reviews, see e.g., \cite{2009SSRv..149..119N} and \cite{2018SSRv..214...45M}).
However, it remains to show a definitive piece of observational evidence demonstrating the existence of sausage waves in coronal structures (see e.g., section 4 in \cite{2019ApJ...874...87S}).

Recently, due to high spatial and temporal observations provided by ground-based instruments, the direct evidence of sausage modes in the lower solar atmosphere have been reported \citep{2014A&A...563A..12D, 2015ApJ...806..132G,2016ApJ...817...44F}.  
\cite{2014A&A...563A..12D} investigated the perturbation relations between cross-sectional areas and total intensities in a sunspot and two pores. They detected the sausage modes with periods from 4 to 65 minutes. By analyzing the ratio between different periodic values, they further revealed that the oscillation modes can be considered as standing waves. 
Multiple-channel observations are employed to investigate upward propagating sausage waves from the lower photosphere to the upper photosphere/lower chromosphere \citep{2015ApJ...806..132G}. According to the energies carried by sausage modes, the authors found that the wave energies decrease substantially with heights and inferred that the energies could be released into the surrounding chromospheric plasma. \cite{2016ApJ...817...44F} inferred that the sausage modes detected in two photospheric pores are standing harmonics with strong reflections at the transition region.  \cite{2018ApJ...857...28K} presents evidence of surface and body modes in photospheric pores supporting sausage waves, the frequency ranges between 2 and 12 mHz.

The purpose of this study is to analyze the perturbations in both areas and intensities within
a nearly circular sunspot umbra observed in two different passbands, 
thereby looking for evidence of sausage waves propagating from photospheric to chromospheric heights.
This paper is structured as follows:  Section~\ref{sec_data} describes our observations and data reduction. Our results are presented in Section~\ref{result}. Finally, Section~\ref{disc} discusses and concludes this study.  
\begin{figure*}[t]
	\centering
	\includegraphics[width=\textwidth]{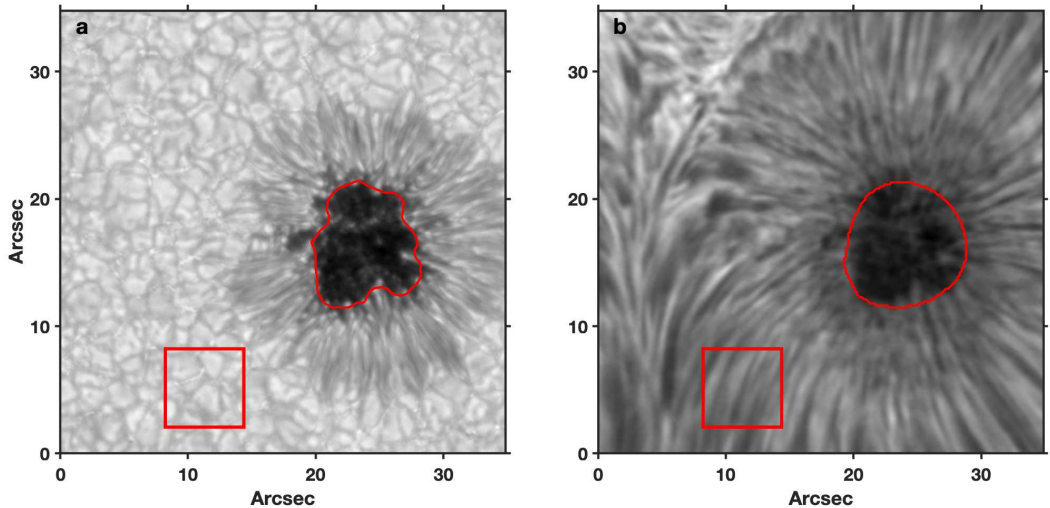}
	\caption{Two images observed by the NVST instrument in the TiO (panel a) and H$_\alpha$  (panel b) channels at 01:32:29 UT 2013 August 6. Two red contours indicate the umbral edge identified, respectively. The two boxes mark the quiet regions that were used to normalize the intensities in the TiO and H$_\alpha$ images, respectively. This step ensures that the disturbance introduced by atmosphere of the Earth is removed. }
	\label{fig1}
\end{figure*}

\begin{figure*}[t]
	\centering
	\includegraphics[width=\textwidth]{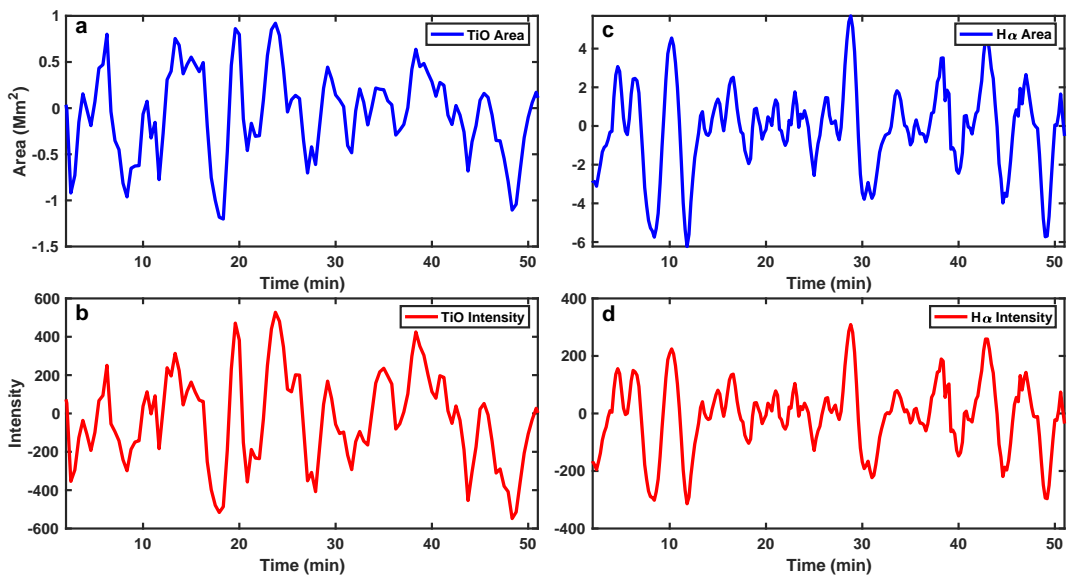}
	\caption{Variations of the umbral areas (a) and intensities (b) obtained from the TiO data set.  (c) and (d) are the counterparts obtained in the H$_\alpha$ data set.}
	\label{fig2}
\end{figure*}

\section{Observations and Data reduction}
\label{sec_data}

In this study, we focus on active region NOAA 11809, its heliocentric co-ordinates was (95\arcsec, 20\arcsec) on 2013 August 6. Two co-spatial high-resolution image sequences were acquired by the New Vacuum Solar Telescope~\citep[NVST;][]{2014RAA....14..705L}. The H$_\alpha$ (6562.8 $\pm$ 0.25 \AA) and TiO (7058 $\pm$ 10~\AA) channels were used to capture the chromospheric and photospheric images between 01:32 and 02:26 UT. The sampling interval was 25 s and 12 s for TiO and H$_\alpha$ images, respectively. Each pixel in TiO images corresponded to 0\arcsec.041 on the sun; whereas in H$_\alpha$ images a pixel has an angular with of 0\arcsec.165. Each sequence was co-aligned with sub-pixel accuracy with a user-defined registration algorithm \citep{Feng:2012hk}. The TiO images were aligned with the H$_\alpha$ images after scaling and rotation. The formation height of the TiO channel is located at the lower part of the photosphere, and H$_\alpha$ at the upper part of the chromosphere.  

Figure \ref{fig1} shows the field-of-view provided by the TiO and H$_\alpha$ channels of NVST. The selected sunspot was close to the disk center, and its shape was a nearly circular. 
We could consider this sunspot as a magnetic waveguide filled with stratified solar atmosphere. Therefore, this sunspot is an ideal model to investigate the periodic oscillations between area and intensity variations. 

The area of sunspot umbra was detected by a thresholding method. To remove the effect of interference of strong penumbra emission, we used a boxcar with 25$\times$25 pixels for the TiO data and 19$\times$19 pixels for the H$_\alpha$ data to smooth the contours. Figure \ref{fig1}a and b shows detection of the area of the umbra. The emission intensity was summed up within the contour in each image and were normalized by the average intensity sampled at quiet sun region (Red box in Figure \ref{fig1}). This step ensures that the effect of atmospheric seeing was removed. Figure \ref{fig2} shows the variation of the areas and normalized intensities measured in both TiO and H$_\alpha$ data. Here, the global trend of each time-vary curve was removed by 12 minute moving average to highlight the oscillation signatures. 

Although,  the atmosphere seeing effect is removed by normalization as stated in the previous paragraph, we also further cross-validate the signal with space-borne instrument, the Atmospheric Imaging Assembly (AIA) onboard the Solar Dynamic Observatory (SDO). So we extracted the umbra area and emission intensity with 1700 \AA{} data. The results are presented in the Appendix (see Figure \ref{sdo_lc}). We obtained very similar result with space-borne instrument, but the noise level is stronger than the NVST high-resolution data, so we only present the analysis with NVST channels thereafter.

The temporal variations of umbral area and total intensity were analyzed by synchrosqueezing transform~\citep[SST;][]{Daubechies:2011bg}. SST is a novel time-frequency (TF) analysis method that precisely not only represent the power spectra of an oscillation signal, but also extract the intrinsic modes of the signal like empirical mode decomposition \citep{1998RSPSA.454..903E} but even higher accuracy \cite{Daubechies:2011bg}. It is a derivation of continuous wavelet transform (CWT) empowered by a spectral reassignment algorithm that compensates the spreading inherent from CWT. The reassignment algorithm uses the phase information of CWT and concentrates on the spectral energy only along the frequency direction, therefore, it sharpens the wavelet spectrum, and in the meanwhile preserves the time resolution of the signal. Because SST concentrates on oscillatory modes into a narrow region in the TF plane and the energy concentration processing is invertible, this allows us to isolate and extract the energy ridge. In the TF plane, the local maxima are defined as the energy ridge, which indicates the frequency distributions of an intrinsic mode. SST uses a penalized forward-backward greedy algorithm to extract the energy ridges~\citep{Daubechies:2011bg}. This algorithm optionally constrains jumps in frequency with a penalty that is proportional to the square of the distance between frequency bins. With this approach, the transform not only provides a perfect inversion, but also gives a stable the reconstruction.  To non-stationary signals, SST not only provides spectral representation but also reconstructs its intrinsic modes as done in empirical mode decomposition. The phase relation between area and intensity oscillations is calculated to constrain the MHD wave mode, therefore, their intrinsic modes were reconstructed to calculate the oscillation phase, see Figures~\ref{tio} and \ref{ha}. 
To SDO data, the spectra and the modes  are presented in the Appendix (see Figures \ref{1700}).

\section{Result}
\label{result}

\begin{figure*}[t]
	\centering
	\includegraphics[width=\textwidth]{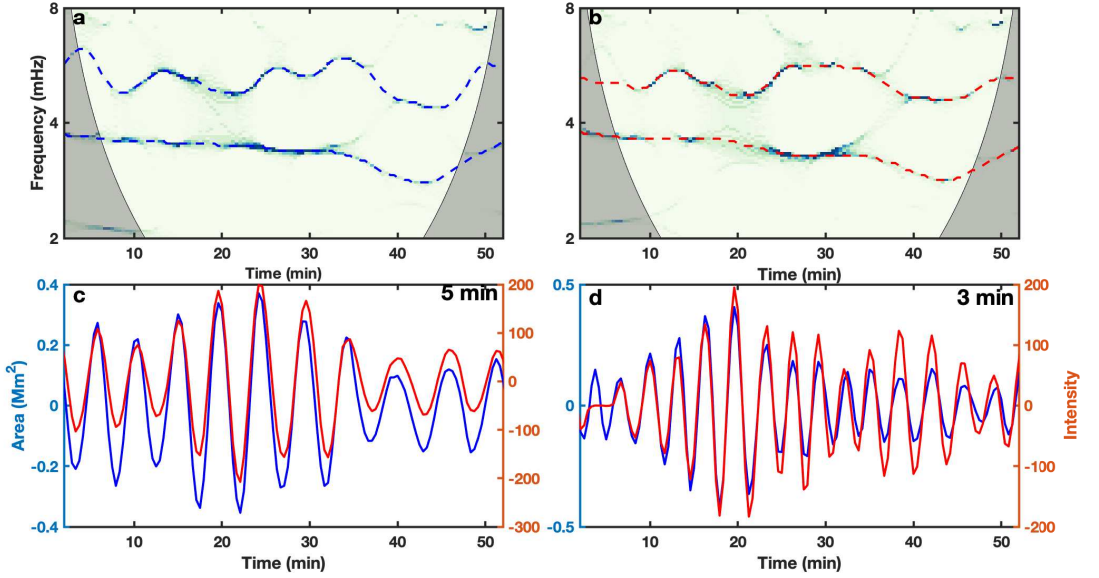}
	\caption{SST power spectra of the area (a) and intensity (b) perturbations as plotted in Figure \ref{fig2}a and \ref{fig2}b. The thin dashed lines indicate the positions of the energy ridges that represent the local maxima. (d) and (c) are reconstructed signal in the three and five minute bands, respectively. The area is plotted with blue lines, and the intensity with red lines. This analysis was done with the TiO data set. }
	\label{tio}
\end{figure*}

\begin{figure*}[t]
	\centering
	\includegraphics[width=\textwidth]{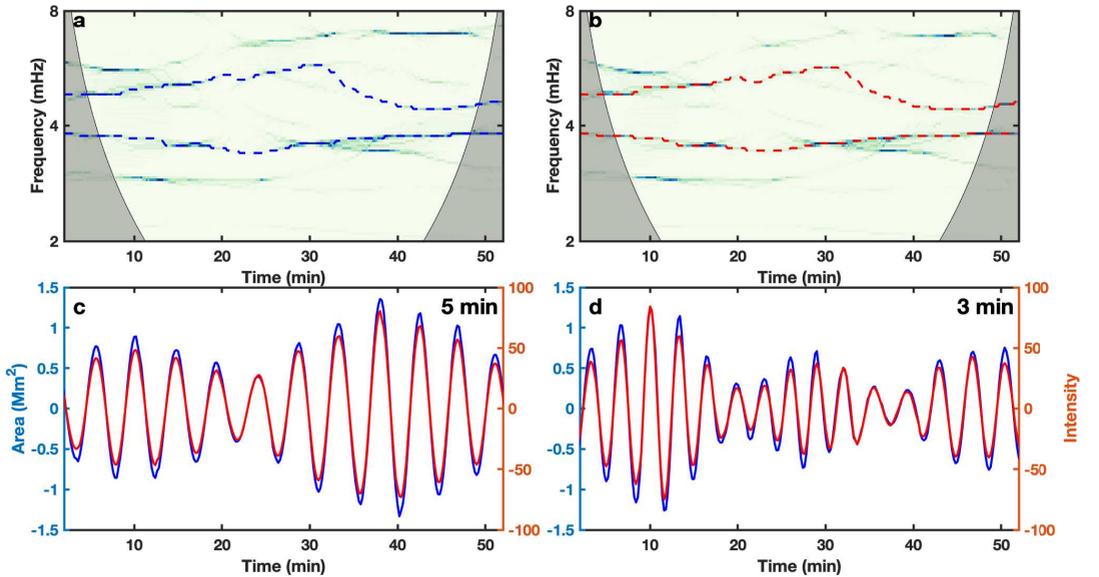}
	\caption{Same as Figure \ref{tio} but with the H$_\alpha$ data set.}
	\label{ha}
\end{figure*}

The NVST TiO channel records the dynamics at the photospheric height, we use this data set to analysis the umbra oscillation therein. Figure~\ref{tio}a and b show power spectra of the umbral area and intensity perturbations plotted in Figure \ref{fig2}.  We could see that significant oscillation power were detected at 3--4 mHz (five minute bandpass) and 5--6 mHz (three minute bandpass). The SST power spectrum of both the area variation (Figure~\ref{tio}a) and intensity (Figure~\ref{tio}b) were very similar to each other. 

We construct the narrowband oscillatory signal of both umbra area and emission intensity at three (Figure~\ref{tio}d) and five minute bandpass (Figure~\ref{tio}c), the reconstructed signal could reveal the phase relationship between the area and intensity variation. We could see the five minute oscillation, the umbra area oscillated in phase with the emission intensity, this is a signal of propagating slow mode wave, see the simulation in \cite{2016ApJS..224...30Y}.

Similar analysis was done to the NVST H$_\alpha$ channel, which observe the dynamics at the chromosphere of the sunspot. We detect similar oscillatory signals and phase relationship in this channel. The detected periods in both area and intensity variations for NVST channels are summarized in Table \ref{IMT_period}.

\begin{table}[!ht]
	\centering  
	\caption{Mean and standard deviation of three and five periods}
	\begin{tabular}{lllll}
		\hline\hline
		&                                                    &TiO              &H$_\alpha$         & Lag Time\\
		\hline
		\multirow{2}{*}{Five minutes band}& Area		  	&4.94 $\pm$ 0.39			&4.55$\pm$ 0.18	 & 180 $\pm$ 12 s		\\
		&Intensity	    &4.95 $\pm$ 0.36			&4.53$\pm$0.15   &   \\ 
		\multirow{2}{*}{Three minutes band}&Area		     &2.89 $\pm$ 0.39			 &3.36 $\pm$0.26  & $168\pm12$ s		 	\\
		&Intensity	     	&3.21$\pm$ 0.23	          	&3.20$\pm$0.25   &  \\
		\hline
	\end{tabular}
	\label{IMT_period}
\end{table}

To study the propagation feature of the umbral wave between two layers observed by the NVST H$_\alpha$ and TiO channels, we combined these two channels and measured the lag time between them. 
The cross-correlation was calculated between the area variation measured at the TiO channel and that in the H$_\alpha$. The lag time for the five minute band is about $168\pm12$~s; whereas the counterpart in the three minute band was about $180\pm12$s.

\section{Conclusions and Discussions}
\label{disc}

In this study, we used the NVST multi-channel observation and analysed the oscillatory signals in sunspot AR 11809. This sunspot was a unipolar sunspot, which has an ideal magnetic and thermal structure and could act as an effective waveguide for MHD waves. The area and intensity variations of this sunspot were extracted in both the TiO and H$_\alpha$ channels. 

In the area and intensity variations of the TiO channel, we detected two oscillatory signals at three minutes and five minutes band. We found that at both bands, the intensity variation oscillated in phase with the area variation. This is the feature of a propagating longitudinal wave \citep{2016ApJS..224...30Y}. Similar results were found in the H$_\alpha$ channel.

If we combined two channels at TiO and H$_\alpha$, which record the dynamics at the photospheric and chromospheric heights, respectively, we obtained a lag time of about $180\pm12$ s. If we assume that the height between the TiO and H$_\alpha$ channels is about 1500 $\pm$ 250 km \citep{Avrett_2008}, the phase speeds for the propagating wave was about 8.9 $\pm$ 0.18 km$\cdot$s$^{-1}$ for five minute band. For the three minute band, we obtained a propagation speed at 8.3 $\pm$ 0.18 km$\cdot$s$^{-1}$. 

For a propagation slow mode wave, the oscillation period is found at three minutes and five minute bands. The propagation speed are about 8--9 km$\cdot$ s$^{-1}$, so the local temperature is about 2800--3500 K. This temperature is smaller than temperature minimum of an empirical sunspot model \citep{Avrett_2008}.  However, we should note this is lower limit estimations, since we did not consider the projection effect. Such a study could also be used as an effective tool to probe the temperature of a sunspot.      

\begin{acknowledgements}
The authors gratefully acknowledge the anonymous referee for his/her critical reading and invaluable comments and suggestions.
Song Feng is supported by the Joint Funds of the National Natural Science Foundation of China (U1931107) and the Key Applied Basic Research Program of Yunnan Province (2018FA035). Ding Yuan is supported by the National Natural Science Foundation of China (11803005, 11911530690) and Shenzhen Technology Project (JCYJ20180306172239618). We thank the Open Research Program (KLSA202007 and KLSA201814) of Key Laboratory of Solar Activity of National Astronomical Observatory of China.
We thank the NVST team for their high-resolution observations. 
\end{acknowledgements}

\section*{Appendix}
\begin{figure*}[!h]
	\centering
	\includegraphics[width=\textwidth]{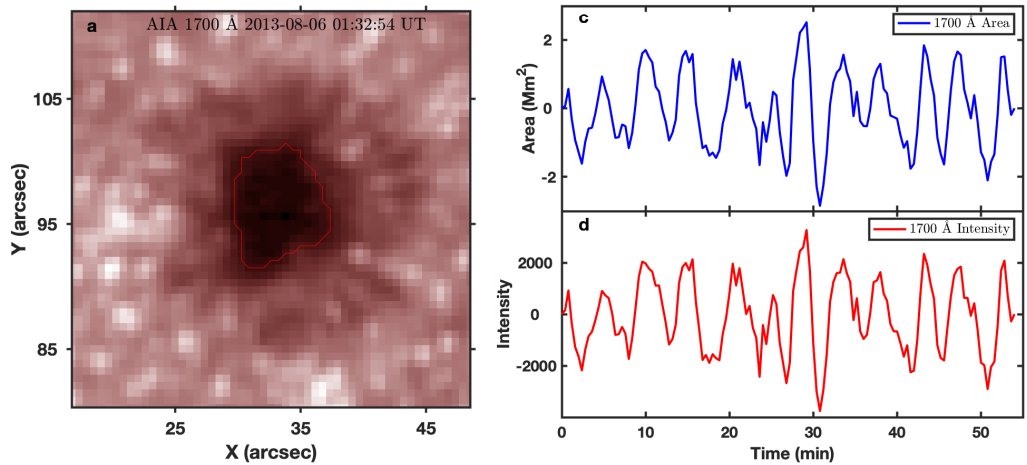}
	\caption{Same as Figures \ref{fig1} and \ref{fig2} but with the AIA 1700 \AA{} data set.}
	\label{sdo_lc}
\end{figure*}
\begin{figure*}[!h]
	\centering
	\includegraphics[width=\textwidth]{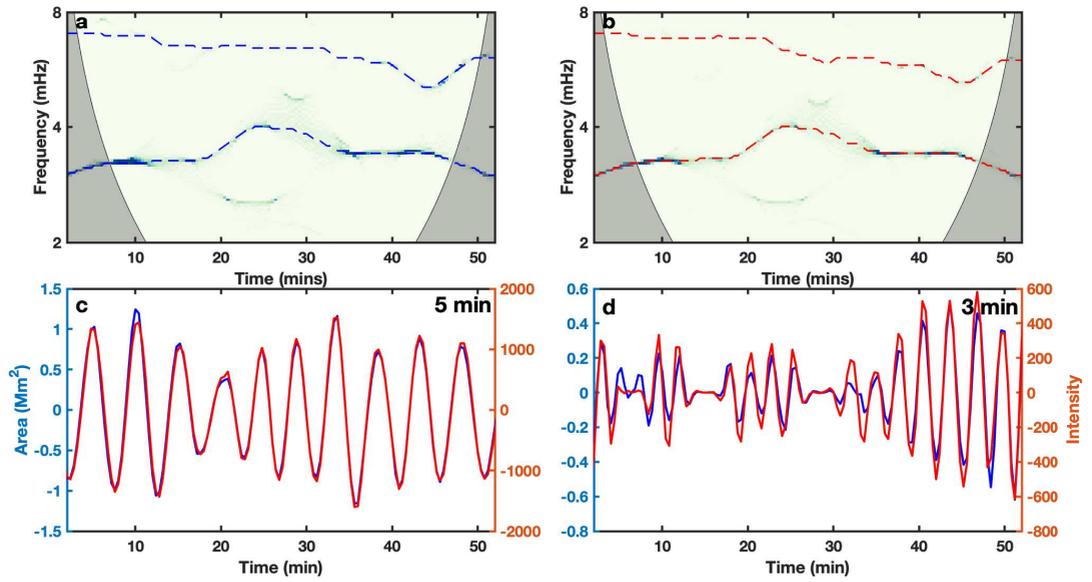}
	\caption{Same as Figure \ref{tio} but with the AIA 1700 \AA{} data set.}
	\label{1700}
\end{figure*}


\begin{thebibliography}{32}
	\providecommand\natexlab[1]{#1}
	\providecommand\JournalTitle[1]{#1}
	
	\bibitem[Avrett \& Loeser(2008)]{Avrett_2008}
	Avrett, E.~H., \& Loeser, R. 2008, The Astrophysical Journal Supplement Series,
	175, 229
	
	\bibitem[{Bai} {et~al.}(2013)]{2013SoPh..282..405B}
	{Bai}, X.~Y., {Deng}, Y.~Y., \& {Su}, J.~T. 2013, Solar Physics, 282, 405
	
	\bibitem[{Bai} {et~al.}(2014)]{2014MNRAS.445...49B}
	{Bai}, X.~Y., {Deng}, Y.~Y., {Teng}, F., {et~al.} 2014, \mnras, 445, 49
	
	\bibitem[{Chen} {et~al.}(2018)]{2018ApJ...868....5C}
	{Chen}, S.-X., {Li}, B., {Shi}, M., \& {Yu}, H. 2018, \apj, 868, 5
	
	\bibitem[Chen {et~al.}(2015)]{2015SoPh..290.2231C}
	Chen, S.-X., Li, B., Xia, L.-D., \& Yu, H. 2015, Solar Physics, 290, 2231
	
	\bibitem[Daubechies {et~al.}(2011)]{Daubechies:2011bg}
	Daubechies, I., Lu, J., \& Wu, H.-T. 2011, Applied and Computational Harmonic
	Analysis, 30, 243
	
	\bibitem[Dorotovic {et~al.}(2014)]{2014A&A...563A..12D}
	Dorotovic, I., Erd{\'e}lyi, R., Freij, N., Karlovsk{\'{y}}, V., \& M{\'a}rquez,
	I. 2014, Astronomy {\&} Astrophysics, 563, A12
	
	\bibitem[{Edwin} \& {Roberts}(1983)]{1983SoPh...88..179E}
	{Edwin}, P.~M., \& {Roberts}, B. 1983, \solphys, 88, 179
	
	\bibitem[Feng {et~al.}(2012)]{Feng:2012hk}
	Feng, S., Deng, L., Shu, G., {et~al.} 2012, in 2012 IEEE Fifth International
	Conference on Advanced Computational Intelligence (ICACI (IEEE), 626
	
	\bibitem[Freij {et~al.}(2016)]{2016ApJ...817...44F}
	Freij, N., Dorotovic, I., Morton, R.~J., {et~al.} 2016, The Astrophysical
	Journal, 817, 44
	
	\bibitem[Goossens {et~al.}(2013)]{2013ApJ...768..191G}
	Goossens, M., Van~Doorsselaere, T., Soler, R., \& Verth, G. 2013, The
	Astrophysical Journal, 768, 191
	
	\bibitem[Grant {et~al.}(2015)]{2015ApJ...806..132G}
	Grant, S. D.~T., Jess, D.~B., Moreels, M.~G., {et~al.} 2015, The Astrophysical
	Journal, 806, 132
	
	\bibitem[Huang {et~al.}(1998)]{1998RSPSA.454..903E}
	Huang, N.~E., Shen, Z., Long, S.~R., {et~al.} 1998, in Royal Society of London
	Proceedings Series A, NASA Goddard Space Flight Center, 903
	
	\bibitem[Inglis {et~al.}(2009)]{2009A&A...503..569I}
	Inglis, A.~R., Van~Doorsselaere, T., Brady, C.~S., \& Nakariakov, V.~M. 2009,
	Astronomy \& Astrophysics, 503, 569
	
	\bibitem[Jess {et~al.}(2013)]{2013ApJ...779..168J}
	Jess, D.~B., Reznikova, V.~E., Van~Doorsselaere, T., Keys, P.~H., \& Mackay,
	D.~H. 2013, The Astrophysical Journal, 779, 168
	
	\bibitem[Keys {et~al.}(2018)]{2018ApJ...857...28K}
	Keys, P.~H., Morton, R.~J., Jess, D.~B., {et~al.} 2018, The Astrophysical
	Journal, 857, 28
	
	\bibitem[Li {et~al.}(2014)]{2014A&A...568A..31L}
	Li, B., Chen, S.-X., Xia, L.-D., \& Yu, H. 2014, Astronomy {\&} Astrophysics,
	568, A31
	
	\bibitem[Liu {et~al.}(2014)]{2014RAA....14..705L}
	Liu, Z., Xu, J., Gu, B.-Z., {et~al.} 2014, Research in Astronomy and
	Astrophysics, 14, 705
	
	\bibitem[{McLaughlin} {et~al.}(2018)]{2018SSRv..214...45M}
	{McLaughlin}, J.~A., {Nakariakov}, V.~M., {Dominique}, M., {Jel{\'\i}nek}, P.,
	\& {Takasao}, S. 2018, \ssr, 214, 45
	
	\bibitem[Moreels {et~al.}(2015)]{2015A&A...579A..73M}
	Moreels, M.~G., Freij, N., Erd{\'e}lyi, R., Van~Doorsselaere, T., \& Verth, G.
	2015, Astronomy {\&} Astrophysics, 579, A73
	
	\bibitem[Moreels {et~al.}(2013)]{2013A&A...555A..75M}
	Moreels, M.~G., Goossens, M., \& Van~Doorsselaere, T. 2013, Astronomy {\&}
	Astrophysics, 555, A75
	
	\bibitem[Nakariakov {et~al.}(2012)]{2012ApJ...761..134N}
	Nakariakov, V.~M., Hornsey, C., \& Melnikov, V.~F. 2012, The Astrophysical
	Journal, 761, 134
	
	\bibitem[{Nakariakov} \& {Melnikov}(2009)]{2009SSRv..149..119N}
	{Nakariakov}, V.~M., \& {Melnikov}, V.~F. 2009, \ssr, 149, 119
	
	\bibitem[{Shi} {et~al.}(2019)]{2019ApJ...874...87S}
	{Shi}, M., {Li}, B., {Huang}, Z., \& {Chen}, S.-X. 2019, \apj, 874, 87
	
	\bibitem[Su {et~al.}(2016)]{2016ApJ...816...30S}
	Su, J.~T., Ji, K.~F., Banerjee, D., {et~al.} 2016, The Astrophysical Journal,
	816, 30
	
	\bibitem[Tian {et~al.}(2016)]{2016ApJ...823L..16T}
	Tian, H., Young, P.~R., Reeves, K.~K., {et~al.} 2016, The Astrophysical Journal
	Letters, 823, L16
	
	\bibitem[Vasheghani~Farahani {et~al.}(2014)]{2014ApJ...781...92V}
	Vasheghani~Farahani, S., Hornsey, C., Van~Doorsselaere, T., \& Goossens, M.
	2014, The Astrophysical Journal, 781, 92
	
	\bibitem[Wang {et~al.}(2020)]{2020raa}
	Wang, Z., Feng, S., Deng, L., \& Meng, Y. 2020, Research in Astronomy and
	Astrophysics, 20, 6
	
	\bibitem[Yu {et~al.}(2016)]{2016ApJ...833...51Y}
	Yu, H., Li, B., Chen, S.-X., Xiong, M., \& Guo, M.-Z. 2016, The Astrophysical
	Journal, 833, 51
	
	\bibitem[{Yuan} {et~al.}(2016)]{2016ApJS..224...30Y}
	{Yuan}, D., {Su}, J., {Jiao}, F., \& {Walsh}, R.~W. 2016, Astrophysical Journal
	Supplement Series, 224, 30
	
	\bibitem[Yuan {et~al.}(2015)]{2015ApJ...807...98Y}
	Yuan, D., Van~Doorsselaere, T., Banerjee, D., \& Antolin, P. 2015, The
	Astrophysical Journal, 807, 98
	
	\bibitem[Yuan \& Walsh(2016)]{2016A&A...594A.101Y}
	Yuan, D., \& Walsh, R.~W. 2016, Astronomy {\&} Astrophysics, 594, A101
	
\end{thebibliography}
\end{document}